\documentclass{elsart3p}

\usepackage{epsfig}

\usepackage{amssymb}

\usepackage{xspace}

\def\1{\footnotemark[1]}
\def\and{\& }

\def\gcm2{g/cm$^2$\xspace}

\begin{document}

\begin{frontmatter}

\title{Cherenkov Flashes and Fluorescence Flares on Telescopes: New lights on UHECR Spectroscopy while unveiling Neutrinos Astronomy}

\author{D. Fargion$\;\;$}
\ead{daniele.fargion@roma1.infn.it}
\author{P. Oliva$\;\;$}
\ead{pietro.oliva@roma1.infn.it}
\author{F. Massa$\;\;$}
\author{G. Moreno}

\address{Physics Department and INFN, University  Sapienza, Rome, Italy}

\begin{abstract}
 Multi-GeV and TeVs gamma sources are currently observed by their
Cherenkov flashes on Telescopes (as Magic, Hess and Veritas),
looking vertically up into sky. These detectors while pointing
horizontally should reveal also the fluorescence flare tails of
nearby down-going airshowers. Such airshowers, born at higher (tens
km) altitudes, are growing and extending up to lowest atmospheres
(EeVs) or up to higher (few km) quotas (PeVs). These fluorescence
signals extend the Cherenkov telescopes to a much higher Cosmic Ray
Spectroscopy. Viceversa, as it has been foreseen \cite{Fargion2005}
and only recently observed, the opposite takes place. Fluorescence
Telescopes made for UHECR detection (as AUGER ones) may be blazed by
inclined Cherenkov lights: less energetic, but frequent (PeVs) CR
are expected to be often detected. Nearly dozens blazing Cherenkov
at EeV should be already found each year in AUGER, possibly in
hybrid mode (FD-SD, Fluorescence and/or Surface Detector). Many more
CR events (tens or hundred of thousands) at PeVs energies should be
blaze Cherenkov lights each year on the AUGER Fluorescence
Telescopes. Their UV filter may partially hide their signals and
they cannot, unfortunately, be seen yet in any hybrid mode. At these
comparable energy the rarest UHE resonant antineutrino
$\bar{\nu}_e+e$ interactions in air at $\frac{{M_{W}}^2}{2m_e}=6.3$
PeV  energy, offer enhanced $W^-$ Neutrino Astronomy showering at
air horizon, at $\sim90^\circ$, while crossing deep atmosphere
column depth or Earth (Ande) boundaries. However, AUGER FD are
facing opposite way. An additional decay channel rises also (after
resonant neutrino skimming Earth) via their secondary $\tau$ exit in
air, by decay in flight via amplified showering:
$\bar{\nu}_e+e\rightarrow W^-\rightarrow\bar{\nu}_{\tau}+\tau$.
Moreover, expected horizontal UHE GZK neutrinos
$\nu_{\tau}\,\bar{\nu}_{\tau}$ at EeVs energy, powered by guaranteed
cosmogenic GZK \cite{Greisen:1966jv, za66},
$\nu_{\mu}\,\bar{\nu}_{\mu}$ flavor conversions (in cosmic
distances), are also producing penetrating  UHE EeV lepton taus that
could sample, better and deeper than PeVs ones, the Earth skin. Such
almost horizontal and up going tau showers, originated by UHE
astronomical neutrino, may shower and flash by Fluorescence and/or
Cherenkov diffused lights at Auger Sky in a few years (nearly
three). Viceversa,  at Hess, MAGIC and VERITAS Horizons, at tens or
a hundred kilometer distances, the same up going $\tau\,\bar{\tau}$
airshowers  might rise via fluorescence. On axis they might blaze
(rarely) as a Cherenkov flashes below the horizons, possibly
correlated to BL Lac or GRB activity. Also UHE ($1-0.1$ EeV) GZK
$\tau$ showering, can be observed upward once reflected onto clouds.
The geomagnetic splitting may tag the energy as well as the inclined
shower footprint as seen in a recent peculiar event in  AUGER.
Additional stereoscopic detection may define the event origination
distance and its consequent primary composition, extending our
understanding on UHECR composition.
\end{abstract}

\journal{Nuclear Instruments and Methods A}
\end{frontmatter}

\section{Introduction}

\begin{figure}
\centering
\includegraphics[width=7.5cm]{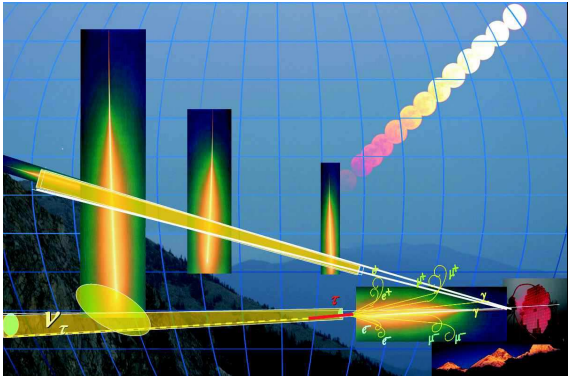}
\caption{The different geometry for a Magic-like telescope searching
at horizon airshower tails. Nearby PeVs CR can be observed by
fluorescence at few km altitude while a more powerful airshower
(EeV), might be observed at lower altitudes because of the larger
slant depth on far edges. Cherenkov airshowers might blaze in the
telescope if in axis. Rarest up-going Tau airshower may escape the
Earth at far horizon, leading to up-going flashes. Guaranteed nearby
CR (PeVs-EeVs) airshower Cherenkov lights, may also be  observed by
reflection from nearby hills or mountains (as in Magic or Veritas),
or from the sea \cite{Fargion07}}\label{fig1}
\end{figure}
\label{intro} Cosmic Rays is a mature, century old  Science. It
still hides its secrets beyond the amazing homogeneity and isotropy
in all energy range. Only photons, our best  neutral courier in
Astronomy, offered up to now a view of the Universe from lowest
(radio) to highest (TeVs) energies. Because of the TeVs-PeVs-EeVs
photon self-interaction with relic IR (Infrared), CBBR (Cosmic Black
Body Radiations) and Radio backgrounds, UHE (Ultra High Energy)
photons are bounded in nearby Universe. Therefore neutral neutrinos
may offer a far insight of CR sources because of their weak
interaction  for we could also reveal the inner core of their
mysterious accelerators.  After a decade, the exciting hopes of a
discover by AGASA and by Hires of (an almost undeflected) UHECR
traces, a long waited new Particle Astronomy have been frustrated by
AUGER results. No UHECR clusterings toward BL Lacs or AGN seems to
arise up now. Even if GZK cut-off (the CBBR opacity to nucleons
above few $10^{19}$ eV flux \cite{Greisen:1966jv, za66}) appeared to
be finally confirmed by Hires \cite{Hires06} and AUGER
\cite{Yamamoto2007}: no Galactic or nearby Universe map within GZK
cut off volume seems to be correlated with these UHECR events. If
homogeneity and isotropy will survive AUGER, Z-Burst model
\cite{Fargion-Mele-Salis99}, linking far cosmic UHE ZeV $\nu$
sources scattering on  $\bar{\nu}$  relic ones, remains the unique
natural option. Otherwise, our Near Universe as Virgo and our Super-
Galactic group and/or plane, must rise soon. Moreover an unexpected
heavy composition of extreme UHECR makes more urgent an independent
UHECR spectroscopy. As well as the discover of UHE $\nu\,\bar{\nu}$
GZK, rare but guaranteed GZK \cite{Greisen:1966jv, za66} secondary
neutrino traces. To this project and to its solution we address in
present paper.
\begin{figure}[]
\centering
\includegraphics[width=7.5cm]{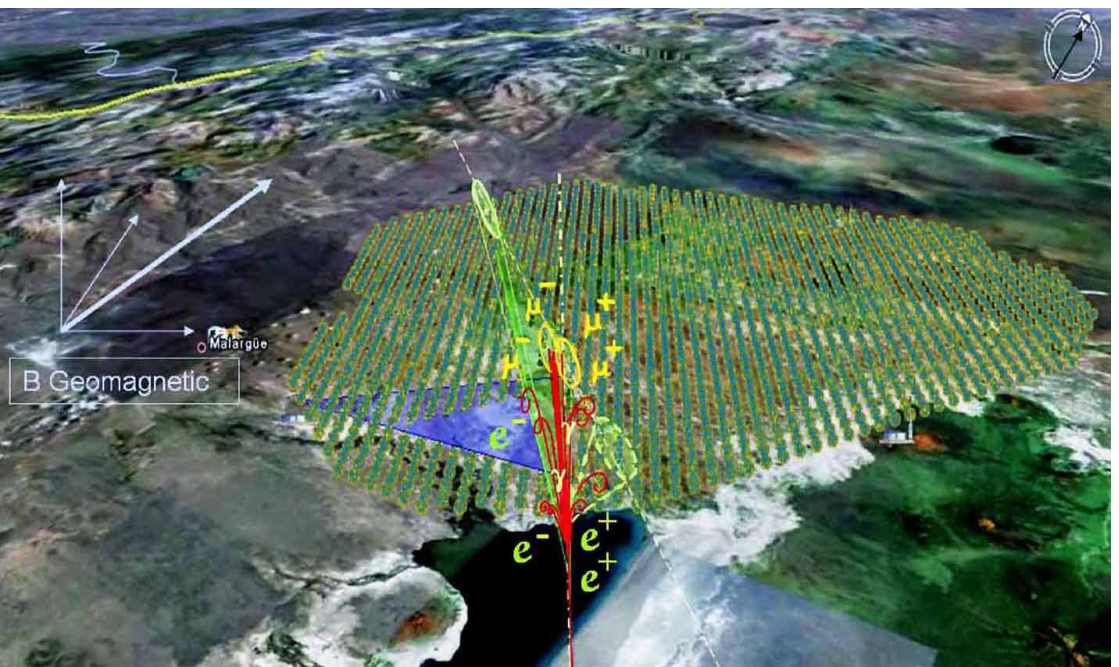}
\caption{The rare inclined UHECR event seen in axis from above. To
this picture derived by an AUGER presentation, we overlapped a
drawing of the airshower components, their bending and the
consequent geomagnetic $\vec{B}_\oplus$ splitting. Note the
$\vec{B}_\oplus$ vector pointing to North and upward respect the
ground. Note also the consequent vertical and lateral charge bundle
separations. Each charge-bundle follow its bend trajectory that
generate its own Cherenkov beam. The comparability between
$\vec{B}_{\oplus\perp}$ and $\vec{B}_{\oplus\parallel}$ field vector
modules, is the cause of a similar angular (lateral-vertical)
deflection. Their projection on the ground is not symmetric at all.
The dashed-ellipse on the right side marks our forecast Cherenkov
spot made by $e^+$ split shower component, undetected (out of a
Cherenkov  reflection on the ground hardly recorded) by Los Leones
FD station which instead detected the main Florescence flare. The
top-left side ellipse, marks the probable Cherenkov spot born by
negative electrons blazing  on Coihueco telescope.}\label{fig2a}
\includegraphics[width=7.5cm]{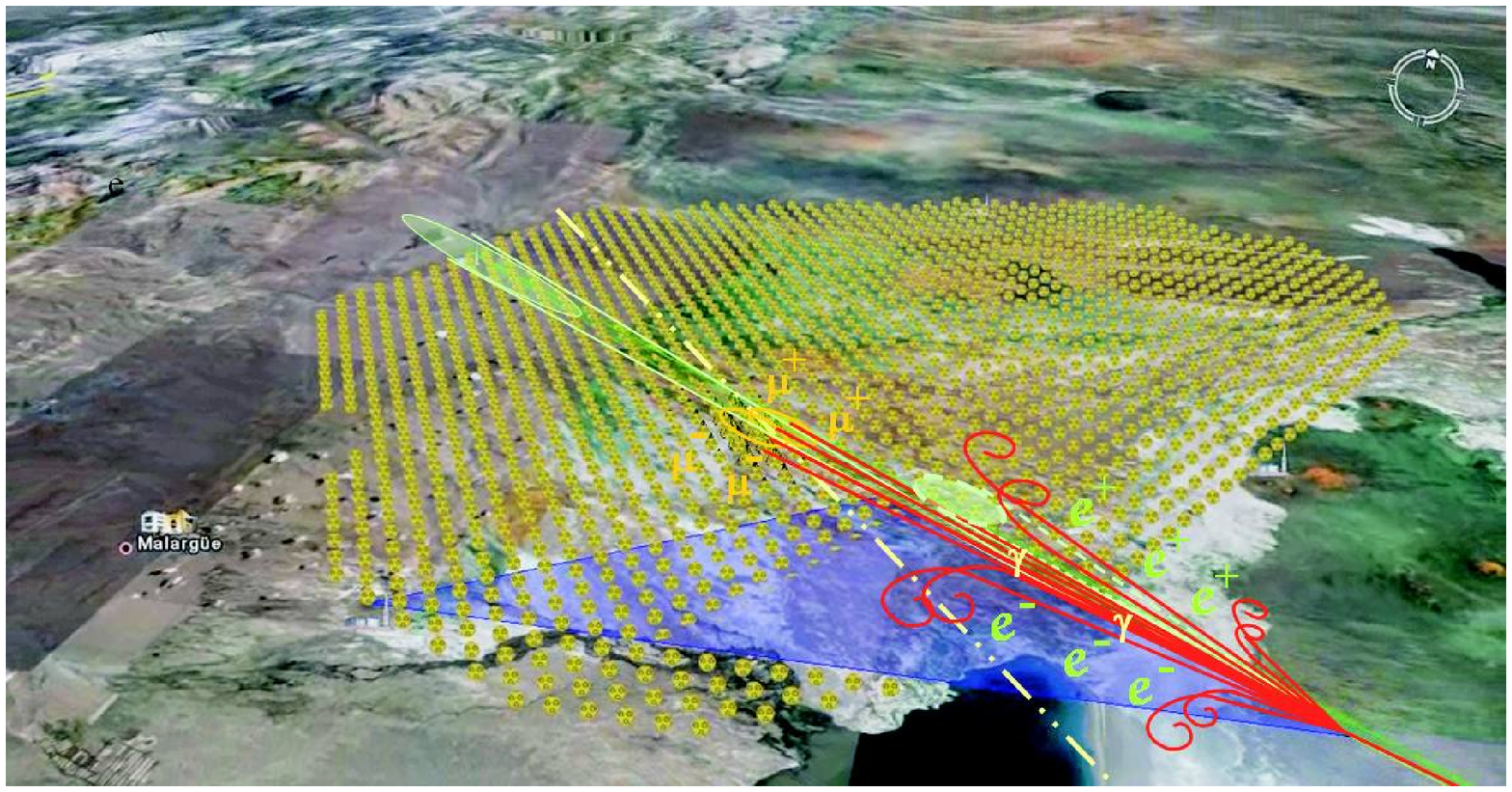}
\caption{The same Auger event seen from another angulation
\cite{Auger07}. The inclined UHECR is reaching the ground on the
center, where the muon pairs are clustering into a twin overlapped
ellipses. The arrival angle is approximated at $80^\circ$ zenith
angle \cite{Auger07}. Our consequent estimated primary energy is
above few EeV (possibly around ten EeV). The  Cherenkov blaze time
on Coihueco $\tau'$, being slightly off axis ($\sim5^\circ$), must
appear much shorter (one or two order of magnitude) than the
Fluorescence duration signal $\simeq\tau$ reaching Los Leones,
because of both the geometric and the relativistic shrinkage,
$\tau'=\tau_o\cdot(1-\beta\cos(\theta))$. The FD, because of lack of
angular resolution in Auger Telescope, might be unable to reveal the
electron pair splitting. The Cherenkov bending angle extends a few
degrees: at Coihueco the airshower beam  is roughly at $5^\circ$
above horizons, while its shower beginning reaches $7^\circ-9^\circ$
angle. Therefore the Cherenkov splitting shape might be marginally
detectable by Coihueco Telescope, by a two-three pixels separation
along its inclined polarized axis.}\label{fig2b}
\end{figure}

\subsection{Fluorescence  flares within Cherenkov Telescope}

We foresee that Cherenkov telescopes while pointing at horizons
($\geq80^\circ$ zenith angles) may observe a truncate image (a
cylinder like) of downward fluorescence airshower, lightening (See
Fig.\ref{fig1}). These flaring views may appear often at few tens km
distance, toward $80^\circ$ angle for tens $PeVs$ (or a hundred km
at $\simeq85^\circ$ for EeVs energies), as often as once a night.
Nearby hills or reflecting sea may disturb the detection. We foresee
that such a discover must occur soon, amplifying MAGIC, VERITAS and
HESS high energy CR yields.

\subsection{Cherenkov blazing photons on Fluorescence Telescopes}

The opposite also take place: Cherenkov photons may hit Fluorescence
Telescopes, even if  most Fluorescence  detectors are masked by UV
filter. Indeed the blazing Cherenkov lights are collimated into a
narrow cone; therefore they are more rarer than Fluorescence
isotropic signals. We may estimate that few hundreds of EeV
airshowers in a year might hit with Cherenkov the AUGER FD. Probably
only a few dozens are well revealed with the FD and by muons on SD,
as well as Cherenkov lights, as it has been foreseen
\cite{Fargion2005}. A very peculiar and pedagogical event has been
shared on line by AUGER collaboration \cite{Auger07}. We shall
analyze that event in the next sections.

\section{Splitting and bending of UHECR at horizons}

The bending of geomagnetic field plays a key role in inclined
airshowers. In usual \emph{vertical} airshowers, the final result is
a conical three-like shape, because the electron pairs avalanche is
spread by random walk in Coulomb scattering at final sea level
altitude. On the contrary, inclined airshower develop at high quota
at low air density, where the Coulomb scattering is negligible. The
Lorentz force may separate better the electron pairs (mostly at GeV
energy) into a twin arc jet-like tails. The same bending takes place
for muons on the ground (recorded by SD). However, because of large
inclined distances and because of muon decay in flight, not to
mention  their larger slant depth, the surviving muon bundles reach
the Earth at higher energy ($\simeq20$ GeV at $80^\circ$) than
electron ones. The muon Lorentz radius is consequently much larger
(nearly $20$ times at $80^\circ$) than electron ones. Therefore
inclined muons suffer less bending  than $e^\pm$ pairs bent in
earlier showering stages, than the muon lobes often overlaps as in
Fig.\ref{fig2a}. Moreover, the geomagnetic $B_{\oplus}$ field is not
(usually) just parallel ($B_{\parallel}$) to the Earth, but it
shares an escaping (or ingoing the Earth) vertical component,
$B_{\perp}$. In particular at South Pole (close to the  AUGER
location), or eventually in the North one, $B_{\perp}$ is comparable
to $B_{\parallel}$. In conclusion the  $B_{\parallel}$ field splits
a downward inclined (horizontal) event, respectively into a twin
upper-down component while $B_{\perp}$ field opens the twin showers
into right-left direction. Both effects usually occurs, defining a
correlated polarized axis. This footprint offers a powerfull tool in
airshower diagnostic. We foresaw and discussed such a splitting role
for Auger in past papers \cite{Fargion2005}.

\section{Inclined Auger Cherenkov-Fluorescence event}

The inclined airshowers experience deep air column depth that filter
their survival probability and intensity at each zenith angle. Such
a tuned filter suppresses the lowest energy showers but leads to a
complex footprint for the others. Indeed the larger the hadronic
airshowers zenith angle is, the higher their main showering
(electron pairs) altitude must be. At highest quota, the air density
is so low, $\rho=\rho_oe^{-\frac{h}{h_o}}$, that the pair energy
threshold increases $E_{min}=20$ MeV $\cdot e^{\frac{h}{2h_o}}$.
Consequently also the Cherenkov  luminosity  decreases
$\frac{dN_{ph}}{dx}\simeq 25e^{-\frac{h}{h_o}}m^{-1}$,  and the
consequent Cherenkov airshower angle shrinkage:
$\theta_{Ch.}=1.4^\circ\cdot e^{-\frac{h}{2h_o}}$, where the air
scale factor is $h_o=7.25$ km. We remind that fluorescence yield is
almost stable up to $20$ km altitude and at higher altitudes it
decays linearly with height. The geomagnetic field, unable to
deflect the UHECR primary, may well deflect its low energy
secondaries (mostly $e^\pm$ and only partially their higher energy
muons $\mu^\pm$). The most inclined the airshower ($85^\circ$ or
above) must rise their main showering lights (Fluorescence or
Cherenkov) at highest altitudes ($\simeq 20-30$ km). Because of the
present Auger solid angle view, extending up to $30^\circ$ from
horizon, such an inclined horizontal shower must be located very far
($\geq 40$ km) edges from the telescopes. At those distances only
highest event might be rarely observed by Fluorescence lights, but
not by Surface Array. And viceversa: extreme inclined events seen by
Cherenkov tanks cannot be observed (because geometry) by FD. At
lower altitudes \emph{only neutrino may shine horizontally} as
discussed below. One of us predicted the inclined
Cherenkov-Fluorescence event in earlier papers \cite{Fargion2005}.
Such a rare inclined pedagogical event (see figure captions
Fig.\ref{fig2a} and Fig.\ref{fig2b}) summarize the splitting in
FD-Cherenkov lights: a high energy airshower has been detected by FD
in Los Leones while being observed by Coihueco Fluorescence
Telescope, via the shower Cherenkov light. The blaze rise from
highest altitudes and its main axis reaches the ground at the Auger
array center where muon bundles hit the tanks. Being muons at $\sim
20$ GeV, they are less bent than the electron pairs in main shower.
The electron components shine in Cherenkov mode, quite inclined as
described in  Fig.\ref{fig2a} and Fig.\ref{fig2b} toward Coihueco
and elsewhere in the array as shown by dashed ellipse;  The exact
calibration of the airshower earlier interaction defines the primary
crossection as well the consequent primary nature: the higher
altitude for heavier nuclei, the lower quota for a nucleon.

\section{Unveiling GZK Neutrino Showering}

\begin{figure}
\centering
\includegraphics[width=8.7cm]{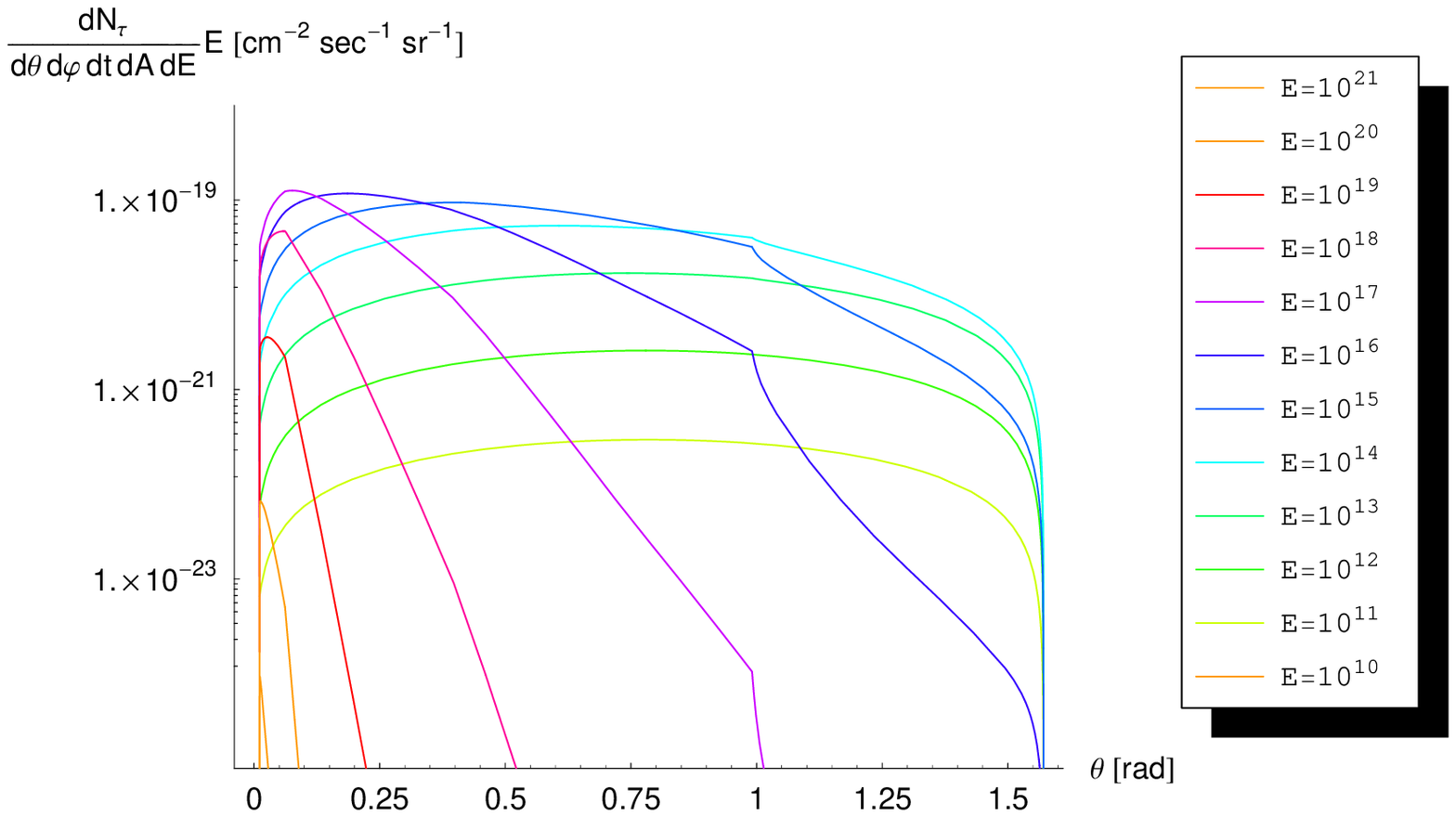}
\caption{Estimate of differential angular number flux of Tau
airshowers at different energies   due to an average (Fermi-like or
Waxman-Bachall like) GZK neutrino fluency around  EeV energy, for
each flavor state, equal to $\Phi_{\nu} = 10 eV cm^{-2}s^{-1}sr^{-1}
$. One may note the negligible absorption at $10^{10}-10^{14}$ eV,
the larger and larger opacity at higher energies toward vertical
directions, the discontinuity at $57^\circ$ due to inner terrestrial
core density step, the surviving tau flux at horizon edges.
\cite{Fargion2004, Fargion2005}; For comparison see \cite{zas05}.
}\label{Fig4}
\includegraphics[width=4.8cm,angle=270]{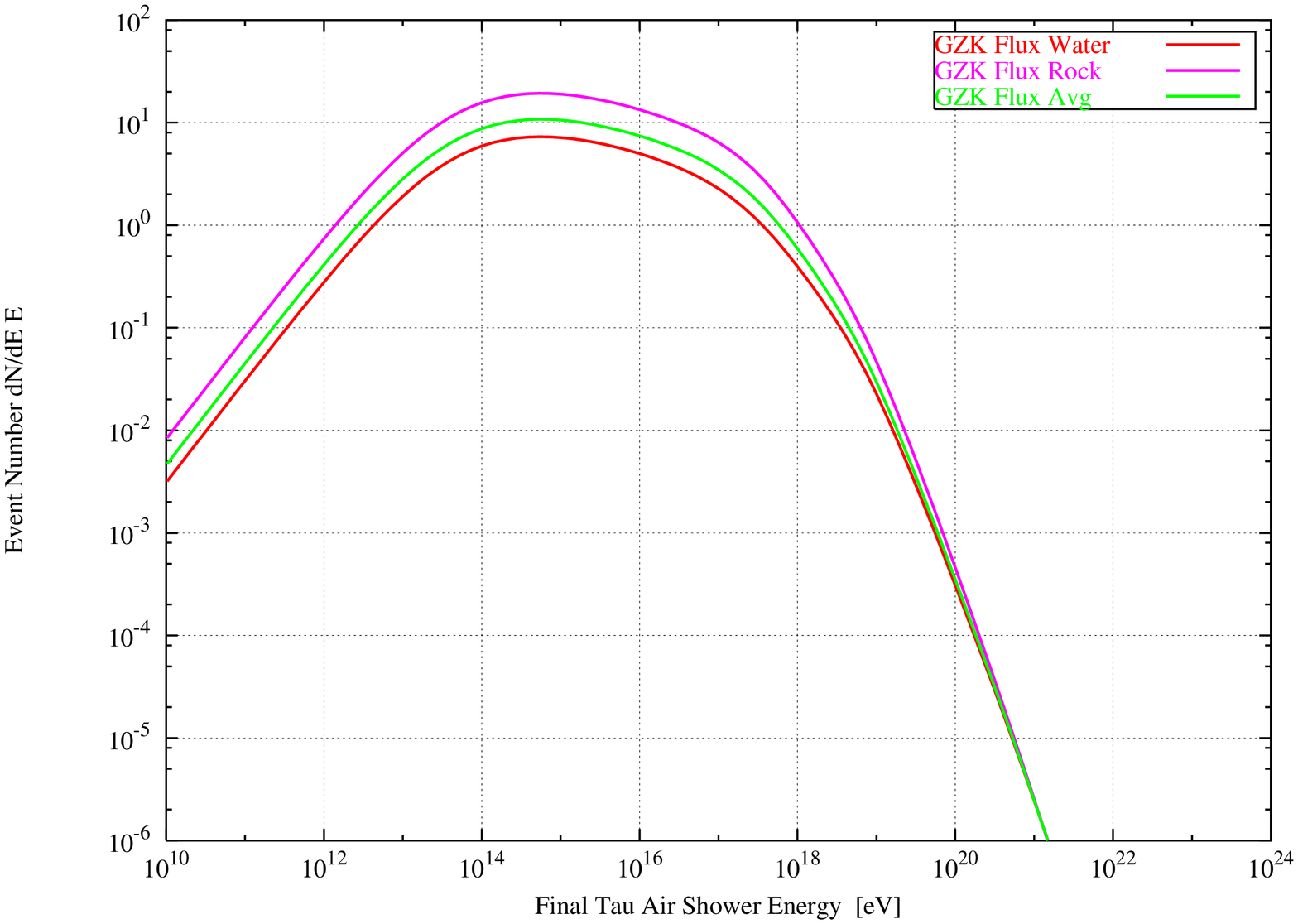}
\caption{Estimate angular integral event rate in three years at
AUGER area as above. As it may note at EeV edges, on rock soil,
where AUGER detection reaches the whole area, there must be an event
in three years from now. The rate maybe increased by Cherenkov
reflection lights on clouds at $0.1-1 EeV$ energy range
\cite{Fargion2007}.}\label{Fig5}
\end{figure}

To estimate a minimal GZK neutrino flux \cite{Greisen:1966jv, za66}
we note that the Auger UHECR at GZK knee  $ E = 3.98  \cdot 10^{19}
eV$ is corresponding to a small fluency ($\Phi_{GZK}\simeq 6.6 eV
\cdot cm^{-2}s^{-1} sr^{-1}$); this value for sake of simplicity may
hint an underlying neutrino GZK flux as large as
($\Phi_{GZK_{\nu}}\simeq 10 eV \cdot cm^{-2}s^{-1} sr^{-1}$ ) for
each flavor state: $\Phi_{\nu_{\tau}+\overline{\nu}_{\tau}}\simeq 20
eV \cdot cm^{-2}s^{-1} sr^{-1}$  for up-going GZK   taus
\cite{Fargion2004, Fargion2007}. This value should be  not too far
from the real one. For simplicity we assumed around EeV energy a
minimal flat ($\propto E^{-2}$) neutrino $\tau$ spectra (the sum of
both two species), comparable with the WB one,  at a constant
fluency. Our results (differential and integral flux) for Auger  are
summarized in Fig. \ref{Fig4},\ref{Fig5}. It is evident that at EeV
in rock matter (as the one in Auger territory),  the expected rate
reach one event in three years. An enhancement, made by  peculiar
Ande screen, may amplify the rate from the West side (at least
doubling  this expected rate).

One  signature  of young Neutrino airshower is its curvature and
time structure: they may indicate the Tau  neutrino origin
\cite{Bertou2002}. However, it is not the unique and most powerful
imprint.  The Auger angular resolution and its limited statistics
will not   allow to reveal any Moon or Solar Shadows, at least in a
decade. The Ande shadow \cite{Fargion1999, Miele06} however is at
least a thousand times larger than the moon one; however  on the
horizons the UHECR rate decreases drastically, nearly three order of
magnitude;  nevertheless the West-East asymmetry would rise around
$\geq 88^\circ$ horizons at Coihueco and Los Leones, as a few
hundred missing or asymmetric events, making meaningful its
detection in one year.   It $must$ be observed soon by tuned trigger
and angular resolution care. Its  discover would be an important
crosscheck of the Auger  experiment.    Within this Ande shadow
horizons, taus might be better born, at least one any three   year,
mostly in FD (Fluorescence Detector) but also in SD (Surface
Detector). In this decade Auger may find up-going Tau in its whole
area  at the rate (see Fig.\ref{Fig4}) of $N_{10^{18}eV}=1.07$ event
each three years; at lower energy,  $N_{3\cdot 10^{17}eV}=  3 $  the
Auger area detection is reduced ($\simeq 0.33$), leading to an
important event rate  $N_{E_{\tau}= 3\cdot 10^{17}eV} \simeq 1.1$.
Because additional events  are unconfined (Horizontal) airshower,
this increases the detection mass  and its discover  rate, almost
doubling (from the West) the expectation rate.  Finally a  possible
discover of FD could be amplified by  final flash  via Cherenkov
reflection on the clouds (see Fig.$4$ in \cite{Fargion2007}). Being
cloudy nights a third or a fourth of the whole year, this time may
be an  occasion to exploit even if Moon arises. In partial
disagreement to some earliest\cite{aramo05}  and most recent  Auger
prospects \cite{Bigas07} requiring one or two $decades$ for a WB
flux GZK neutrino discover, we foresee (in see also
\cite{Fargion2007}) a sooner discover of GZK $\tau$ neutrino
astronomy, possibly within  two-three years from now.  Auger may be
even the first  experiment in the world to detect a tau  natural
flavor regeneration processes. To reach and speed this goal we
suggest to tune the electronic trigger of FD to horizontal
airshowers, to map the UHECR Ande shadows, to insert more Cherenkov
telescopes to the Ande line.

\section{ Conclusions}

Cherenkov and Fluorescence Telescope may enlarge their view and role
tracing both lights. The  wider CR range will  allow a better
spectroscopy at PeV-EeV (knee-ankle) regions and a more detailed
anatomy of UHECR composition. MAGIC, HESS and VERITAS may soon trace
the Fluorescence lights of downward UHECR airshowers. MAGIC and
VERITAS must reveal the Cherenkov reflections also on nearby
Mountains. The recent inclined UHECR event in Auger \cite{Auger07}
clearly foreseen in \cite{Fargion2005} discussed in this paper offer
the first footprint for such rich information derived by muons
bundles, electron pairs showering and splitting into polarized
Cherenkov and Fluorescence traces. Within this novel spectroscopy a
hidden Neutrino Astronomy wait to be finally unveiled
\cite{Fargion2007}.
%%%%%%%%%%%%%%%%%%%%%%%%%%%%%%%%

\end{document}